\documentclass[twocolumn,superscriptaddress]{revtex4}
\usepackage{graphicx}
\usepackage{epsfig}
\usepackage{amsmath}
\usepackage{amsfonts,amsbsy}
\usepackage{amssymb}
\usepackage{hyperref}
\usepackage{bm}
\usepackage{color}
\usepackage{enumerate}
\usepackage{verbatim}

\def\be{\begin{equation}}
\def\ee{\end{equation}}
\def\bea{\begin{eqnarray}}
\def\eea{\end{eqnarray}}

\begin{document}
\title{Characteristics of nonequilibrium evolution near the phase boundary}
\author{Xiaobing Li}
\affiliation{Key Laboratory of Quark and Lepton Physics (MOE) and Institute of Particle Physics, Central China Normal University, Wuhan 430079, China}
\affiliation{School of Physics and Electronic Engineering, Hubei University of Arts and Sciences, Xiangyang 441053, China}
\author{Yuming Zhong}
\affiliation{Key Laboratory of Quark and Lepton Physics (MOE) and Institute of Particle Physics, Central China Normal University, Wuhan 430079, China}
\author{Ranran Guo}
\affiliation{Key Laboratory of Quark and Lepton Physics (MOE) and Institute of Particle Physics, Central China Normal University, Wuhan 430079, China}
\author{Mingmei Xu}
\email{xumm@ccnu.edu.cn}
\affiliation{Key Laboratory of Quark and Lepton Physics (MOE) and Institute of Particle Physics, Central China Normal University, Wuhan 430079, China}
\author{Yu Zhou}
\affiliation{Department of Mathematics, University of California, Los Angeles, California 90095, USA}
\author{Jinghua Fu}
\affiliation{Key Laboratory of Quark and Lepton Physics (MOE) and Institute of Particle Physics, Central China Normal University, Wuhan 430079, China}
\author{Yuanfang Wu}
\email{wuyf@ccnu.edu.cn}
\affiliation{Key Laboratory of Quark and Lepton Physics (MOE) and Institute of Particle Physics, Central China Normal University, Wuhan 430079, China}
\date{\today}
\begin{abstract}
Using the single-spin flipping dynamics, we study the nonequilibrium evolution near the entire phase boundary of the 3D Ising model, and find that the average of relaxation time (RT) near the first-order phase transition line (1st-PTL) is significantly larger than that near the critical point (CP). As the system size increases, the average of RT near the 1st-PTL increases at a higher power compared to that near the CP. We further show that RT near the 1st-PTL is not only non-self-averaging, but actually self-diverging: relative variance of RT increases with system size. The presence of coexisting and metastable states results in a substantial increase in randomness near the 1st-PTL, and therefore  makes the equilibrium more difficult to achieve.
\end{abstract}

\maketitle

\section{Introduction}

The exploration of nonequilibrium evolution is essential for understanding of phase transition (PT). Usually, PT refers to first-order phase transition and second-order phase transition. Within phase plane, these two types of PTs form a phase boundary comprising first-order phase transition line (1st-PTL) and critical end point (CP), and extending to a crossover region. This type of phase boundary manifests in diverse systems such as the liquid-gas transition~\cite{nuclear-liquid-gas}, magnetic transition~\cite{magnetic}, metal-insulator transition~\cite{slow-first-order-1}, quark deconfinement, and chiral phase transition in quantum chromodynamics (QCD)~\cite{QCD,high-cumulants,STAR-cumulants}, and so on. As one approaches the phase boundary, the properties of the matter will change significantly.

For the CP, static (equilibrium) and dynamic (nonequilibrium) properties have  been both extensively studied and are well-known~\cite{Newman,Cardy}. For instance, in the thermodynamic limit, the correlation length $\xi$ diverges at the critical temperature $T_{\rm c}$, and the static critical exponents typically take fractional values. 

Meanwhile, in the vicinity of $T_{\rm c}$, the relaxation rate decreases with the correlation length, a phenomenon known as critical slowing down~\cite{1977}. This decrease follows a power-law with respect to the correlation length $\xi$ or the characteristic length of the lattice size $L$, governed by the dynamic critical exponent $z$ which describes the dynamical evolution~\cite{1977,MartinPRE}. 
Furthermore, as the system size increases, the relative variance of susceptibility, specific heat and others approach constant values, indicating non-self-averaging behavior at the CP~\cite{non-selfaveraging-1,non-selfaveraging-2,non-selfaveraging-3,physicsA}.

For the first order PT, the static exponents are typically integers~\cite{integer-1,integer-2,ZhangYH,XPan} which are in contrast to the fractional values at the CP. However, the dynamic properties at and near the 1st-PTL have not been fully confirmed and explored yet. 

Initially, exponential slowing down was observed~\cite{supercritical-1, supercritical-2} and referred to as {\it supercritical slowing down}, which resulted from deterioration in the simulation algorithm. By introducing the multicanonical algorithm, this type of slowing down was improved to follow a power-law. Subsequently, slowing down near the boundary of the coexisting region was observed in solutions of the dynamical mean-field theory of Mott-Hubbard transition~\cite{JooPRB} and later in mechanically induced adsorption-stretching transitions~\cite{zhang2017}. Non-self-averaging behavior has only been observed in the isotropic-to-nematic transition in liquid crystals~\cite{nonSA-first-1}.

To fully reveal the dynamic properties near the entire phase boundary, the nonequilibrium evolution is better presented systematically and simultaneously. The dynamical evolution is commonly described by the Langevin equations~\cite{HSong} and various relaxation models, such as the kinetic Ising model~\cite{kinetic-Ising} or models A, B, C, etc., classified by the renormalization group~\cite{1977}. However, the analytical solutions of dynamic equations turn out to be mathematically difficult. Initially, the solutions were limited to the crossover region~\cite{Mukherjee}, later the solutions were extended to include some of the 1st-PTL region~\cite{langevin-sigma}.

Recently, using the single-spin flipping dynamics, individual relaxation processes near the CP of 3D Ising model have been simulated successfully~\cite{XBLi}. The relaxation time (RT) is defined as the number of sweeps required for the order parameter to reach its equilibrium value. The order parameter is demonstrated to approach its equilibrium value exponentially, which aligns with the behavior predicted by the Langevin equation and the Ising model with a mean-field approximation~\cite{Glauber-analytical}. 

To quantify the RT of a sample, the average of RT is introduced. It has been demonstrated that at the critical temperature, the average of RT follows a power-law increase with system size, and the dynamic exponent is consistent with the dynamic model A~\cite{1977,Roth}. This implies that the average of RT well represents the relaxation time defined analytically. Moreover, the third and fourth cumulants of the order parameter on the crossover side oscillate around zero, and then converge to their equilibrium values. The sign of the third cumulants can be negative, consistent with those obtained from dynamical equations~\cite{Mukherjee,Nahrgang}.  These facts suggest that the single-spin flipping dynamics effectively describe the dynamical relaxation near the CP, and therefore provides a new numerical way to explore nonequilibrium evolution.

In this paper, we employ this numerical way to investigate the nonequilibrium evolution near the entire phase boundary of the 3D Ising model. Firstly, to demonstrate the relaxation features, we start from a random initial configuration and present the average of RT on the phase plane for a specific system size of $L=60$. Secondly, to reveal the randomness which causes the relaxation features, we study the self-averaging properties of RT. Thirdly, to see the influence of initial condition, we parallelly provide and discuss the results obtained from a given high-temperature initial state and a polarized initial state. Finally, we conclude with a brief summary.

\section{Characteristics of nonequilibrium evolution}

The CP of the 3D Ising model belongs to the $Z(2)$ symmetry group. Various physical systems exhibit the same universality class as the 3D Ising model. Examples include the liquid-gas transition~\cite{nuclear-liquid-gas}, magnetic transition~\cite{magnetic}, quark deconfinement, and chiral phase transition in QCD~\cite{Stephanov,chiral-Ising,k4-non}. 

The 3D Ising model considers a three dimensional simple cubic lattice composed of $N=L^3$ spins, where $L$ is called the system size. The total energy of the system with a constant nearest-neighbor interaction $J$ placed in a uniform external field $H$ is 
\begin{equation}
E_{\lbrace s_{i}\rbrace}=-J\sum_{\langle ij\rangle}s_{i}s_{j}-H\sum_{i=1}^N s_{i}, \quad s_{i}=\pm1.
\end{equation}
The per-spin magnetization is
\begin{equation}
m=\frac{1}{N} \sum_{i=1}^N s_{i}.
\end{equation} 
It serves as the order parameter of the continuous phase transition at the critical temperature $T_{\rm c}=4.51$~\cite{tc-Ising}, and below $T_{\rm c}$, there is a 1st-PTL at $H=0$~\cite{XLi}.

 
Single-spin flipping dynamics, e.g. Metropolis algorithm~\cite{Metropolis}, as a local dynamics of Glauber type~\cite{kinetic-Ising}, is suitable for studying nonequilibrium evolution~\cite{Metropolis-NE1,Metropolis-NE2}. Starting from an initial configuration, Metropolis algorithm flips one single spin at each step. Whether a spin flips depends on the acceptance probability $A({\pmb u}\rightarrow {\pmb v})$, which is given by 
\begin{equation}
A({\pmb u}\rightarrow {\pmb v})=\left\{\begin{array}{ll}
e^{-(E_{\pmb v}-E_{\pmb u})/k_{\rm B}T}&\text{if $E_{\pmb v}-E_{\pmb u}>0$},\\1&\text{otherwise.}\end{array}\right .
\end{equation}
${\pmb u}$ and ${\pmb v}$ represent the state of the system before and after flipping this spin. If $A({\pmb u}\rightarrow {\pmb v})=1$, the spin is flipped. If $A({\pmb u}\rightarrow {\pmb v})<1$, a random number $r$ ($0<r<1$) is generated. If $A({\pmb u}\rightarrow {\pmb v})>r$, the spin is flipped, otherwise, the spin keeps its original state. The testing of one single spin is called a Monte Carlo step. When $N$ Monte Carlo steps are completed, every spin in the lattice has been tested for flipping and \textit{one sweep} is completed. In this way, the configuration of the system is updated once a sweep. After evolving enough sweeps, the magnetization approaches a steady value and the system reaches equilibrium. 

Relaxation time (RT) of an individual evolution process is the number of sweeps required for magnetization to reach a stable value~\cite{XBLi}. This definition is in general suitable for all kinds of relaxation processes in reality and for the entire phase space. To quantify the RT of a sample, the average of RT is introduced as~\cite{XBLi}, 
\begin{equation}
\bar\tau=\frac{1}{n}\sum_{i=1}^{n} \tau^{i},
\end{equation}
where $n$ is the total number of evolution processes,  $\tau^{i}$ is RT of the $i^{\rm th}$ process. $\bar\tau$ is equivalent to the relaxation time defined analytically~\cite{XBLi}.

Figure~1 presents a contour plot of the average of RT on the $T$-$H$ phase plane for a fixed system size of $L = 60$, starting from random initial configurations. The color scheme ranges from white to red to black, representing the average of RT values that span from less than a hundred to over four thousand. The phase boundary is depicted by the line $H=0$. In regions far from the phase boundary, the color appears light, indicating a small average of RT. However, a dark-red point emerges around $T_{\rm c}=4.51$, indicating a large average of RT, which is consistent with the phenomenon of critical slowing down as expected.

\begin{figure}[tb]
\centering
\includegraphics[width=0.45\textwidth]{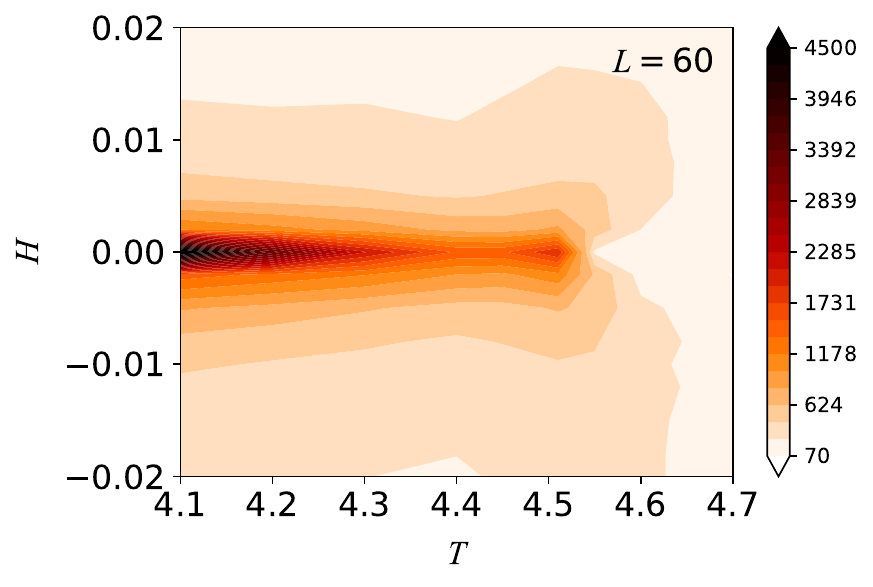}
\caption{A contour plot of the average of RT on the phase plane of the 3D Ising model for $L=60$. The statistics are 10,000 evolution processes for each bin. }
\end{figure}

Along the 1st-PTL, the color becomes progressively darker as the temperature decreases, eventually turning completely black when the temperature drops below 4.2. Simultaneously, on the low-temperature side, specifically along the direction of the external field, the color transitions rapidly to black near $H=0$. These observations indicate that the relaxation process near the 1st-PTL is considerably slower than that near the CP, and can be characterized as {\it ultra-slowing down}.

Ultra-slowing down indicates a higher level of uncertainty and randomness. This can be attributed to the more complex structure of the free energy at the 1st-PT compared to the CP. The 1st-PT represents a transition between distinct internal states, where both upward and downward magnetized phases coexist along the transition line~\cite{Xu-symmetry}. In the equilibrium state, the system can be either in an upward or downward magnetized state with equal probability, effectively doubling the number of possible states. Near the 1st-PTL, some of these possible states manifest as metastable states~\cite{JooPRB,Binder}. The presence of coexisting and metastable states significantly increases the instability, uncertainty, and randomness.  Consequently, achieving equilibrium at the 1st-PT is a challenging task. 

\begin{figure*}[t]
\centering
\includegraphics[width=0.9\textwidth]{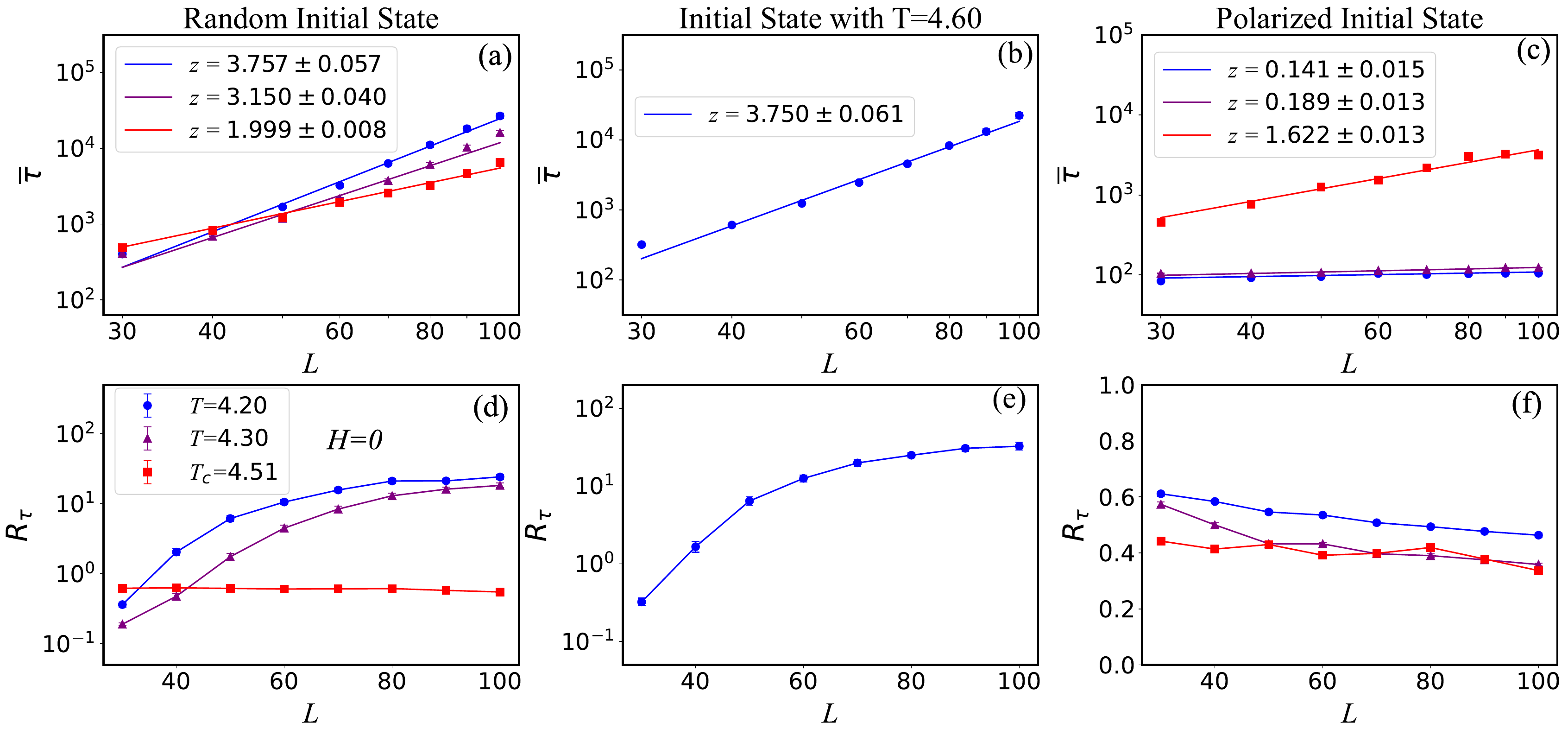}
\caption{The finite size scaling of the average of RT for three different initial states: random initial state (a), equilibrium state at $T=4.60$ (b), and polarized initial state (c). $R_\tau$ as a function of $L$ for each of these cases, denoted as (d), (e), and (f) respectively. The statistics are 20,000 evolution processes for each blue point in Figs. 2(a) and 2(b), 10,000 evolution processes for each point of others.}
\end{figure*}

To show the system size dependence of a nonequilibrium evolution, Fig.~2(a) presents a double-logarithmic plot illustrating the variation of the average of RT with system size for three different temperatures at the boundary. The data points corresponding to critical temperature $T_{\rm c}=4.51$ are denoted by red squares, while the data points at the 1st-PTL with temperature below $T_{\rm c}$ are represented by blue circles and purple triangles, respectively. The error bars indicate only the statistical errors. 

For each of three given temperature, the data points can be well fitted by a straight line. This shows the average of RT increases in a power-law with system size, i.e., 
\begin{equation}
\bar{\tau}\propto L^z,
\end{equation}
at both CP and the 1st-PTL. The exponent of power law $z$, which corresponds to the slope of the fitting line, represents the dynamic exponent. The values of $z$ are provided in the legend of Fig.~2(a). 

The critical dynamic exponent $z=1.999\pm 0.008$ is obtained by fitting from $L=20$ to $L=100$, which has higher statistics and is more precise than the previously reported value of $z=2.06\pm 0.03$~\cite{XBLi}. Both values are consistent with the dynamic model A~\cite{1977,Roth}. The value of dynamic exponent $z$ at the 1st-PTL are larger than that at the CP. This implies that the average of RT increases more rapidly with system size at the 1st-PTL than that at the CP. 

The longer average relaxation time means a larger increase in randomness. To demonstrate the change of randomness of the RT along the entire phase boundary, we examine self-averaging property of RT. Self-averaging refers to the behavior of the relative variance of an observable $X$ as the system size increases. It is defined as follows~\cite{non-selfaveraging-1}:
\begin{equation}
 R_X=\frac{\overline{ X^2}-\overline{ X}^2}{\overline{X}^2}, 
 \end{equation}
where bar represents the average over the entire sample.  If $R_X$ tends to zero, $X$ is self-averaging; if $R_X$  increases, $X$ is self-diverging. In case of self-averaging, the fluctuation of $X$ diminishes as the system size increases, and the average of $X$ converges to the same value. Self-diverging means divergent fluctuations of $X$ as the system size increases.

At the phase boundary, the relative variance of $\tau$ is plotted against the system size for three different temperatures in Fig.~2(d). The trend of $R_\tau$  is obviously different for each of temperature cases.  

At the critical temperature $T_{\rm c}=4.51$, $R_\tau$ remains almost constant as the system size increases. In other words, the variance or width of the RT distribution increases with the system size at the same rate as the
average of RT, indicating non-self-averaging behavior of RT, consistent with other observables~\cite{non-selfaveraging-1}.  

At two lower temperatures $T=4.30$ and $T=4.20$, $R_\tau$ increases with system size, indicating self-diverging behavior at the 1st-PTL, similar to what is observed in liquid crystals~\cite{nonSA-first-1}. As the system size increases, the variance of RT increases more rapidly than the average of RT, and the distribution of RT become wider and flatter. This phenomenon suggests that randomness is significantly amplified with system size, leading to an abnormally increased variance. Such an extremely broad distribution of RT is a feature of the relaxation near the 1st-PTL. 

To explore the behavior away from the boundary, the average of RT for three external fields at a specific temperature $T=4.2$ are plotted against the system size in a double-logarithmic scales in Fig.~3(a). The external fields are $H=10^{-5}$ (red crosses), $H=0.001$ (blue triangles), and $H=0.02$ (green squares). Figure~3(a) illustrates that for each external field, as the system size increases, the average of RT also increases and can be fitted by a line, similar to the case shown in Fig.~2(a). Consequently, away from the boundary, the average of RT also exhibits a power-law relationship with system size.

\begin{figure}[ht]
\centering
\includegraphics[width=0.5\textwidth]{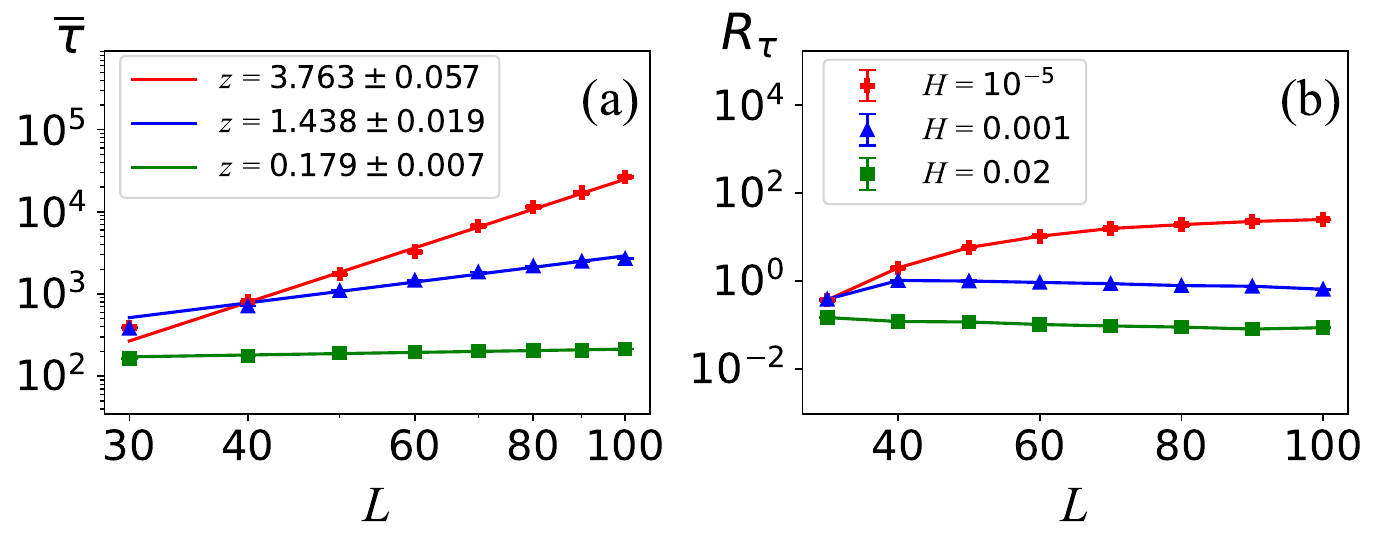}
\caption{(a) The finite size scaling of the average of RT for three different values of the external magnetic field at $T=4.20$. (b) The relative variance of RT as a function of system size for each of the aforementioned cases. The statistics are 20,000 evolution processes per point.}
\end{figure}

The power $z$ for $H=10^{-5}$ as presented in the legend of Fig.~3(a) is nearly the same as that observed at the 1st-PTL shown in Fig.~2(a). However, as one moves away from the 1st-PTL along the direction of the external magnetic field, $z$ gradually decreases. This reduction can be attributed to the alignment tendency of spins with the external magnetic field. As the external field becomes non-zero, the spins tend to align in the direction of the field, thereby reducing the overall randomness in the system.

Furthermore, similar to the case presented in Fig.~3(a), the relative variance $R_\tau$ is plotted against the system size for three external fields in Fig.~3(b). It is observed that self-diverging behavior is only prominent at $H=10^{-5}$, which is the nearest to the 1st-PTL. For larger external fields, such as $H=0.001$ and $H=0.02$, $R_\tau$ decreases slowly as $L$ increases, indicating self-averaging of RT. As the deviation from the 1st-PTL becomes further, the self-diverging behavior of RT becomes less pronounced. Therefore, self-diverging behavior is observed not only precisely at the 1st-PTL but also in its close vicinity, while far from the phase boundary, RT exhibits self-averaging behavior.

 
As demonstrated in Ref.~\cite{XBLi}, the initial configuration has a significant impact on the RT. A random initial configuration represents a completely disordered state, i.e., $|m|=0$, corresponding to a point in the region of spinodal instability~\cite{Xu-symmetry, Randrup-2012}. Nonequilibrium evolution from a very high temperature state (equivalent to a random initial configuration) to the phase boundary can occur spontaneously. To provide a comprehensive analysis, we also include the results for initial states with a temperature of $T=4.6$ and a polarized configuration. 


The initial configuration of Figs.~2(b) and 2(e) corresponds to an equilibrium configuration at $T=4.6$, while the final state temperature is $T=4.2$. Both configurations lie on the boundary $H=0$. It is evident that the initial state with $T=4.6$ is less random than, but still very close to, the random initial configuration. The average of RT shown in Fig.~2(b) also exhibits an increase with system size and can be fitted with a straight line, similar to Fig.~2(a), but with a slightly smaller $z$. Therefore, the divergence of average of RT is not as rapid as in Fig.~2(a) due to the reduced randomness. Simultaneously, the relative variance of RT, as shown in Fig.~2(e), also increases with system size, indicating the occurrence of self-diverging RT.

Figures~2(c) and 2(f) present the average and the relative variance of RT from a polarized initial configuration. In Fig.~2(c), the average of RT increases with system size for each temperature and can be fitted with a line. The slope $z$ at the critical temperature is the largest but smaller than that in Fig.~2(a). The slopes $z$ for the other two temperatures decrease significantly compared to Fig.~2(a). Additionally, Fig.~2(f) demonstrates that the relative variance of RT decreases with system size for all three temperatures, indicating that RT exhibits self-averaging behavior in this case.

As we know, a polarized initial configuration is equivalent to an equilibrium state at a very low temperature. This ordered state possesses a similar structure to the equilibrium states at the 1st-PTL. Consequently, the evolution from a polarized initial configuration to the states at the 1st-PTL is  facile, resulting in a substantial reduction in the average of RT. The self-diverging behavior observed in Fig.~2(d) is completely absent. However, it should be noted that the structure of the polarized configuration still remains distinct from that of the CP. The average of RT towards the CP still diverges.

\section{Summary and discussion}

In summary, using the single-spin flipping dynamics, we simulate the nonequilibrium evolution near the entire phase boundary of the 3D Ising model.

Starting from random initial configurations, the contour plot of average of RT $\bar{\tau} $ in the phase plane for a fixed system size is presented. 
The red area around the CP indicates clearly the known critical slowing down, and black band around the 1st-PTL implies the ultra-slowing down, where relaxation process lasts much longer than that near the CP. The white area of crossover region shows the time of relaxation is negligible. Therefore, the equilibrium is readily achieved in the crossover region, more challenging to attain at the CP, and exceedingly difficult to reach at the 1st-PTL.

We also demonstrate the system size dependence of average RT $\bar{\tau} $ at phase boundary ($H=0$). With the increase of the system size, $\bar{\tau} $ increase in a power-law at both the CP and the 1st-PTL. The dynamic exponent at the CP is consistent with that of dynamic model A. The dynamic exponent at the 1st-PTL is larger than that at the CP. This confirms much longer $\bar{\tau} $ at the 1st-PTL.

Longer RT means an increase of randomness. To examine randomness, the relative variance of RT at the phase boundary are presented. As the system size increases, the relative variance of RT increases at the 1st-PTL, and keeps to a constant at the CP. So RT exhibits a self-diverging behavior at the 1st-PTL, in contrast to the non-self-averaging at the CP. The randomness at the 1st-PTL is enlarged significantly. This increased randomness is attributed to the presence of coexistence states, which greatly expand the number of possible states and consequently lead to a very broad distribution of RT. This characteristic of relaxation at the 1st-PTL distinguishes itself from that observed at the CP.

Parallelly, the power-law behavior of  $\bar{\tau} $ and self-averaging property of RT are presented at different external fields, or near the phase boundary. 
The power-law of $\bar{\tau} $ holds near the phase boundary, and
the value of dynamic exponent decreases away from the phase boundary.
Meanwhile, the self-averaging property of RT also changes from self-diverging to self-averaging away from the phase boundary.

These results confirm that the power-law behavior of  $\bar{\tau} $ and self-averaging property of RT are directly related to the randomness. Approaching the phase boundary, the randomness also increases due to the appearance of metastable state. The increase of randomness leads to a bigger dynamic exponent and a severer violation of self-averaging.

We also discuss the influence of initial condition. If the evolution starts from a high temperature (above  $T_{\rm c}$) initial state, the randomness is reduced. If starting from a polarized initial state, i.e., very low temperature initial state, although critical slowing down still exists, the ultra-slowing down and self-diverging near the 1st-PTL disappear. This is because the very low temperature initial state is very close to the 1st-PTL, and needs not much time to relax. 

To the end, although the specific value of the dynamic exponent may shift systematically with the model and algorithm employed~\cite{scaling-Fischer}, the properties of nonequilibrium relaxation, such as ultra-slowing down and self-diverging at the 1st-PTL, critical slowing down and non-self-averaging, and power-law behavior with increasing system size, should be common and expected across different dynamic models or equations. These properties of nonequilibrium evolution provide deep insights into the specific phase transition process.


\section*{Acknowledgement}
We are grateful to Dr. Yanhua Zhang for very helpful discussions. This research was funded by the National Key Research and Development Program of China, grant number 2022YFA1604900, and the National Natural Science Foundation of China, grant number 12275102. The numerical simulations have been performed on the GPU cluster in the Nuclear Science Computing Center at Central China Normal University (NSC3). 

\providecommand{\href}[2]{#2}\begingroup\raggedright\endgroup
\end{document}